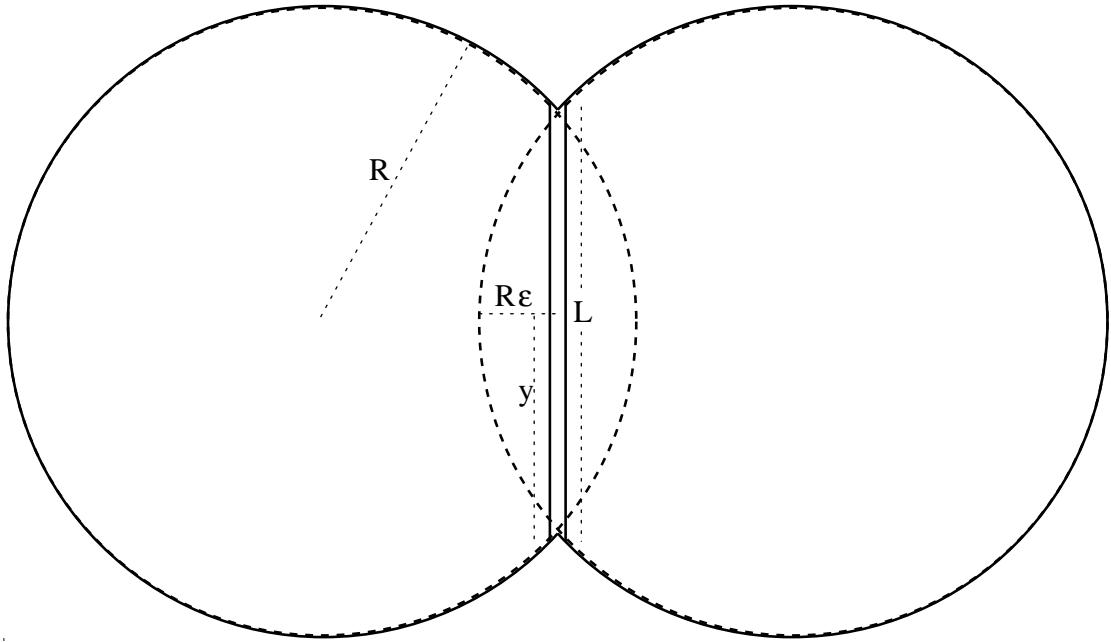

**FIGURE 1.** Sketch of colliding flux tubes in cross section. The solid lines are the tube boundaries. The radius of the isolated tube is $R$, the length of the interaction region is $L$, the $y$ axis is chosen parallel to the contact surface, and $\epsilon$ is the fractional deformation.

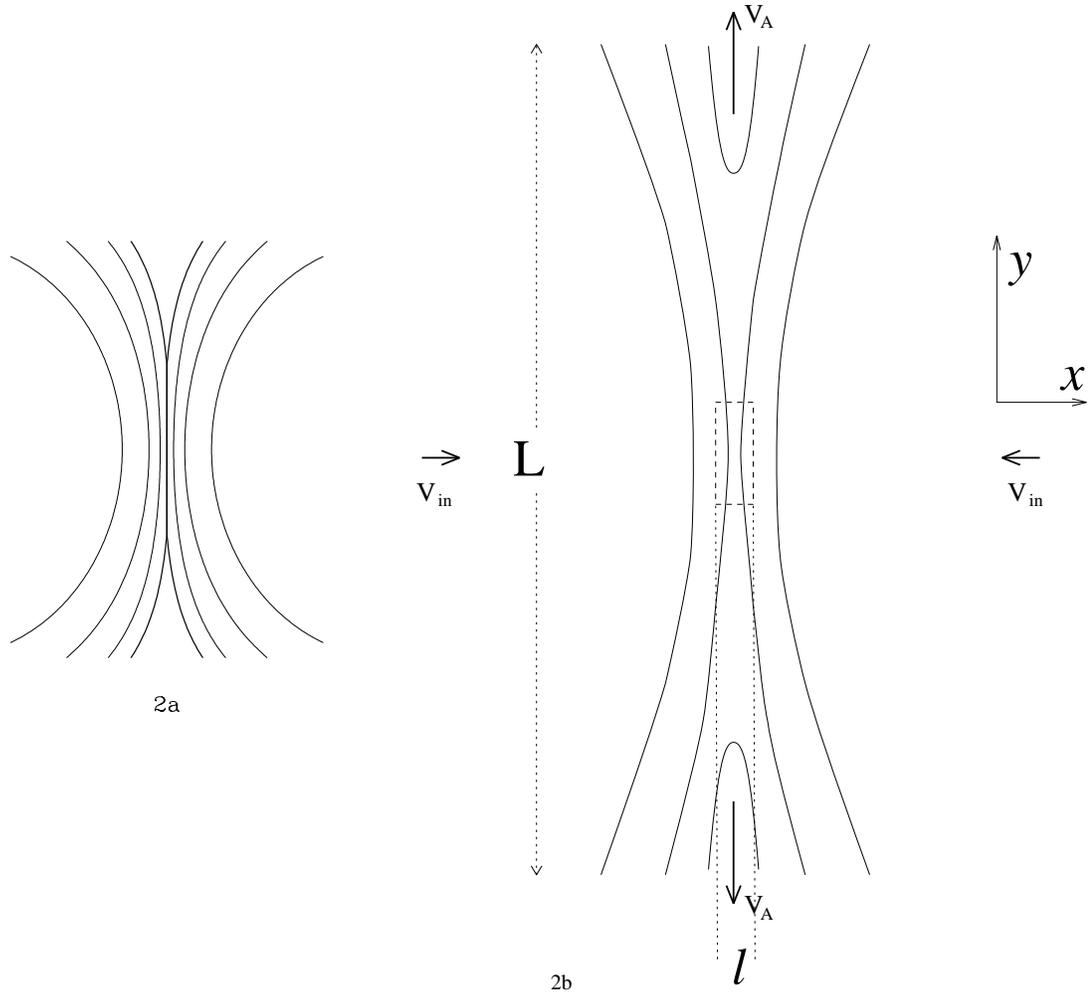

**FIGURE 2.** (a) Sketch of the magnetic flux surfaces of the colliding tubes, projected onto the axial cross section. Reconnection can take place within a thin layer along the contact surface. (b) Sketch of the Parker-Sweet reconnection geometry, which we assert ensues once the configuration shown in Figure (2a) is established by the collision. The $y$ axis is again the contact surface and $v_{in}$ corresponds to the incoming tubes. In slow collisions omly a slight pileup of magnetic flux is sufficient to permit Sweet-Parker reconnection.

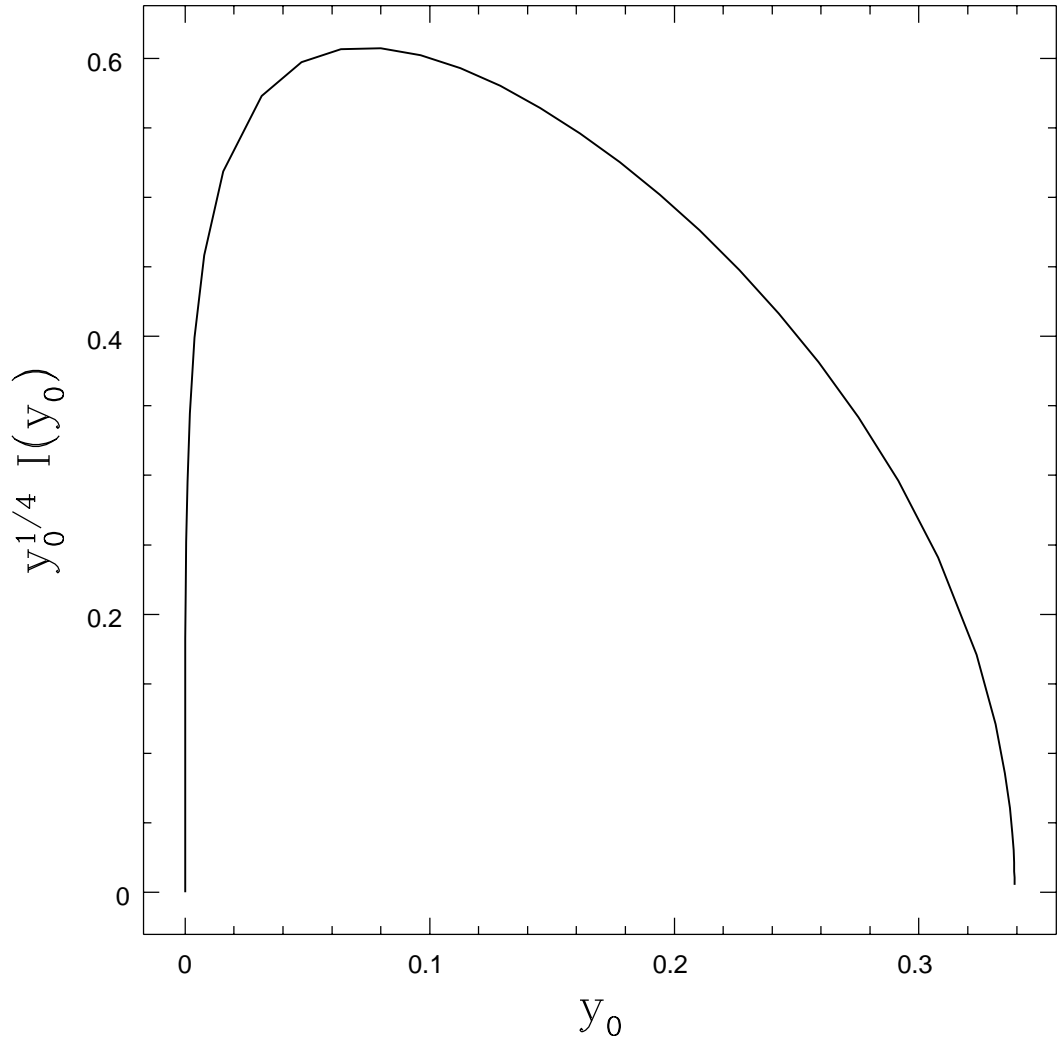

**FIGURE 3.** Plot of the function $y_0^{1/4} I(y_0)$ as a function of $y_0$ over its permitted range. As eqn. (3.12) shows, this function is proportional to the amount of magnetic flux reconnected in a collision.

# MAGNETIC MERGING IN COLLIDING FLUX TUBES

Ellen G. Zweibel[1,2]

Joint Institute for Laboratory Astrophysics, University of Colorado and
National Institute of Standards and Technology, Boulder, CO 80309

and

James E. Rhoads[3]

Department of Astrophysical Sciences, Princeton University, Princeton, NJ 08540

ABSTRACT

We develop an analytical theory of reconnection between colliding, twisted magnetic flux tubes. Our analysis is restricted to direct collisions between parallel tubes and is based on the collision dynamics worked out by Bogdan (1984). We show that there is a range of collision velocities for which neutral point reconnection of the Parker-Sweet type can occur, and a smaller range for which reconnection leads to coalescence. Mean velocities within the solar convection zone are probably significantly greater than the upper limit for coalescence. This suggests that the majority of flux tube collisions do not result in merging, unless the frictional coupling of the tubes to the background flow is extremely strong.

## 1. INTRODUCTION

Observation, theory, and numerical experiment all suggest that magnetic fields become concentrated into sheets and ropes when embedded in gas pressure dominated, turbulent, highly conducting fluids such as stellar convection zones (Galloway & Proctor 1983, Schmitt & Rosner 1983, Hughes and Proctor 1988) and possibly accretion disks (Stella & Rosner 1984, Sakimoto & Coroniti 1989, Schramkowski & Achterberg 1993). These flux tubes move semi- independently of the surrounding fluid: accelerated by buoyancy and magnetic tension forces but coupled to the ambient medium by frictional drag (e.g. Parker 1982, Moreno-Insertis 1986, Choudhuri & Gilman 1987, Chou & Fisher 1989, Fan, Fisher, & DeLuca 1993, DeLuca, Fisher, & Patten 1993).

Under these conditions, magnetic flux tubes are expected to collide with one another. Collisions might result in coalescence, and possible reconnection of the magnetic field, or the tubes might retain their identities and separate. It is clear that the rules which govern collisional interaction, and in particular the reconnection process, play an important role in the evolution of a flux tube dominated magnetic field.

---

[1] Also at Department of Astrophysical, Planetary and Atmospheric Sciences, University of Colorado, Boulder, CO 80309.
[2] I: zweibel@solarz.colorado.edu
[3] I: rhoads@astro.princeton.edu

Bogdan (1984; hereafter B84 ) considered collisions between magnetic flux tubes and developed a model for their merging. He took the flux tubes prior to the collision to be infinite cylinders with both axial and azimuthal components of magnetic field, and he assumed that during the collision the tubes would flatten along a common surface. If the tubes are twisted in the same sense, their azimuthal components in the interaction region are antiparallel, and might be expected to reconnect. B84 treated the reconnection as a $1D$ problem in magnetic diffusion across the common boundary (with the thickness of the boundary taken as a free parameter), and he assumed that the tubes coalesce if the energy of the reconnected magnetic field exceeds the relative kinetic energy of the collision. The amount of reconnected flux is proportional to the magnetic diffusivity $\eta$ which is typically small in situations of interest, so coalescence occurs only for very slow collisions.

The purpose of this paper is to reconsider magnetic reconnection within the framework of B84 's collision model. We argue that reconnection is more likely to proceed through a magnetic $x$−point and to be of the Parker-Sweet type. This determines the width of the diffusion layer and leads to a scaling in which the reconnection rate is proportional to $\eta^{\frac{1}{2}}$. This results in a model in which flux tubes can coalesce even when moving much faster than the critical velocity for coalescence estimated by B84 .

In Section 2 of this paper we review the dynamics of flux tube collisions. In Section 3 we describe the reconnection phase and in Section 4 we estimate the critical velocity below which two flux tubes can merge. Section 5 is a critical review of the assumptions, application of the theory to the solar convection zone, and discussion.

## 2. COLLISIONS BETWEEN MAGNETIC FLUX TUBES

We consider head-on collisions between two identical, parallel magnetic flux tubes. Following B84 , we assume that before contact the tubes are circular cylinders with radius R and that each tube is in magnetostatic equilibrium, with a helical magnetic field $\hat{z}B_z(r) + \hat{\phi}B_\phi(r)$ and gas pressure $P(r)$, where $r$, $z$, and $\phi$ are cylindrical polar coordinates for each tube. We also assume that the relative velocity $v_\infty$ is slow enough that the flows inside and outside the tubes are incompressible. The medium surrounding the tubes is unmagnetized and has uniform density $\rho_e$.

The equation of motion of the tubes prior to contact was solved more than a century ago by Hicks (1879), and the solution is reviewed by B84 . The mass per unit length $\mathcal{M}$ of each tube is increased to an effective mass $\mathcal{M}_{\text{eff}}$ due to the entrainment of ambient fluid: $\mathcal{M}_{\text{eff}} - \mathcal{M}$ is of order $\pi \rho_e R^2$. In stellar interiors the ratio of magnetic to gas pressure inside the tubes is probably rather low, so $\rho_e$ is probably nearly the same as the density $\rho$ in the tubes, and $\frac{\mathcal{M}_{\text{eff}}}{\mathcal{M}}$ should be of order unity.

If the tubes approach each other with speed $\frac{1}{2}v_\infty$ in the center of mass frame, the actual relative velocity $v_c$ at impact tends to be reduced from $v_\infty$ by drag effects, but can also be increased by center of mass motion perpendicular to the direction of approach, and decreased by center of mass motion parallel to the direction of approach. All of these effects are small for $\mathcal{M}_{\text{eff}} \approx \mathcal{M}$, and we will parameterize the collisions by the relative velocity at impact, $v_c$, and refer the reader to eqn. (18) of B84 for the exact relationship between $v_c$, $v_\infty$, and the velocity of the center of mass.

When the tubes collide, they become flattened along a common face. We measure the deformation of each tube by the displacement of a boundary point on the shortest line connecting the $z-$axes of the tubes, and denote this quantity by $R\epsilon(t)$. Then (see Figure 1), the length $L$ of the common face is

$$L = (8\epsilon)^{\frac{1}{2}} R. \tag{2.1}$$

The deformation of the tubes is maintained by an overpressure $\Delta P$ which decelerates the tubes and eventually forces them apart. B84 showed that for small deformations $\Delta P \propto \epsilon B_\phi^2(R)$, and wrote $\Delta P$ as a power series in the distance $\mid y \mid$ from the midpoint of the common boundary (see Figure 1)

$$\Delta P(y) = \frac{\epsilon B_\phi^2(R)}{4\pi} \sum p_n \left(\frac{2y}{L}\right)^{2n}. \tag{2.2}$$

The coefficients $p_n$ were left undetermined by B84, so we treat them as unknowns.

Since the overpressure acts along the length $L$ of the common boundary, eqns. (2.1) and (2.2) show that the restoring force per unit length of tube is proportional to $\epsilon^{\frac{3}{2}}$. The equation of motion for $\epsilon$ is

$$\frac{d^2\epsilon}{dt^2} = -\frac{2}{R\mathcal{M}_{\text{eff}}} \int_0^{\frac{L}{2}} dy \Delta P(y) = -\frac{\epsilon^{\frac{3}{2}} B_\phi^2(R)}{2^{\frac{1}{2}} \pi \mathcal{M}_{\text{eff}}} \sum \frac{p_n}{(2n+1)}. \tag{2.3}$$

The problem up to this point has an intriguing resemblance to the problem of encounters between finite, isotropic elastic bodies (Landau and Lifshitz 1970), which has been solved completely and for which it is also found that the repulsive force between the bodies is proportional to the $\frac{3}{2}$ power of the deformation parameter. We therefore thought that we might be able to find the coefficients $p_n$ by developing an analogy with elasticity theory. However, the force between infinite elastic solid cylinders is found to be proportional to the first power of the deformation parameter, and we were unable to completely cast the MHD equations in the form of elasticity theory (see Appendix), so the analogy proved fruitless and the $p_n$ remain undetermined.

Equation (2.3) has the first integral

$$\frac{1}{2}\left(\frac{d\epsilon}{dt}\right)^2 + \frac{2^{\frac{1}{2}} B_\phi^2(R)}{5\pi\mathcal{M}_{\text{eff}}} \sum \frac{p_n}{(2n+1)} \epsilon^{\frac{5}{2}} = \frac{v_c^2}{8R^2}, \tag{2.4}$$

where the constant of integration follows from the definitions of $\epsilon$ and $v_c$. We now convert eqn. (2.4) to a form that is more convenient for the analysis later in the paper. We introduce the Alfven velocity $v_A$ defined with respect to the azimuthal magnetic field and mass density at the edge of the tube

$$v_A \equiv \frac{B_\phi(R)}{(4\pi\rho(R))^{1/2}}, \tag{2.5}$$

define a parameter $\mu$ which measures the ratio of $\mathcal{M}_{\text{eff}}$ to the mass per unit length the tube would have if its density throughout were equal to the density at the edge

$$\mu \equiv \frac{\mathcal{M}_{\text{eff}}}{\pi \rho(R) R^2}, \tag{2.6}$$

and let the effective Alfven Mach number for the collision be

$$M_{AC}^2 \equiv \frac{\mu v_c^2}{4 v_A^2}. \tag{2.7}$$

We expect that $\mu$ is of order unity and $M_{AC}$ is small for the high $\beta$, slowly moving tubes in the solar convection zone. Finally, we subsume the unknown coefficients $p_n$ into a dimensionless parameter

$$\alpha \equiv \frac{2^{\frac{7}{2}}}{5\pi} \sum \frac{p_n}{(2n+1)}. \tag{2.8}$$

Equation (2.4) then becomes

$$(\frac{d\epsilon}{dt})^2 = \frac{v_c^2}{\mu R^2}(M_{AC}^2 - \alpha \epsilon^{\frac{5}{2}}), \tag{2.9}$$

from which we see that the maximum deformation parameter $\epsilon_{\text{max}}$ is given by

$$\epsilon_{\text{max}} = (\frac{M_{AC}^2}{\alpha})^{\frac{2}{5}}. \tag{2.10}$$

Since $\alpha$ is expected to be of order unity, if we restrict ourselves to small deformations then $M_{AC}$ must be small. The deceleration of the tubes is gentle until $\epsilon$ is close to $\epsilon_{\text{max}}$; $v$ has dropped to $0.5 v_c$ at $\epsilon \sim 0.89 \epsilon_{\text{max}}$, and to $0.1 v_c$ at $\epsilon \sim 0.996 \epsilon_{\text{max}}$.

If the flux tubes are twisted in the same sense, the azimuthal magnetic field components at the deformed interface oppose each other. This suggests the possibility of magnetic reconnection. We estimate the conditions for and amount of reconnection which occurs in the following section.

## 3. MAGNETIC RECONNECTION

### 3.1 Solutions with Flux Buildup

Reconnection between colliding magnetic flux tubes is treated as a diffusion problem by B84 ; the azimuthal magnetic field components in the two tubes are parallel and antiparallel to the contact surface and diffuse in one dimension across a layer which varies in width over time. The reconnection rate derived by B84 is linear in the plasma resistivity $\eta$ and involves an unknown lengthscale: the width of the layer across which the fields diffuse.

We pursue an alternative model of reconnection. It appears likely, considering the time history of the collision, that a magnetic $x$−point forms at the initial point of contact between the tubes. The situation at later times then resembles that sketched in Figure 2a. This configuration is strongly reminiscent of Parker-Sweet reconnection (Parker 1957,

Sweet 1958a,b), and the reconnection is expected to be much faster than $1-D$ diffusion. In this section we find the conditions under which Parker-Sweet reconnection can occur.

We first briefly review the reconnection model (see Parker 1979a for a more complete treatment). Consider the magnetic field shown in Figure 2b. The characteristic length of the system is $L$. Resistivity is important only near the $x$-point, in a layer of width $l$. Fluid is ejected along the $\hat{y}$ axis at the Alfven speed $v_A \equiv \frac{B_y}{(4\pi\rho)^{\frac{1}{2}}}$. The flow is assumed to be incompressible, and therefore the speed $v_{in}$ at which material flows toward the $\hat{y}$ axis is related to $v_A$ by

$$Lv_{in} = lv_A. \tag{3.1}$$

The electric field $E_z$ is determined in the diffusion region (near the $x$-point) by resistivity, and in the ideal region (everywhere else) by inductive effects. Since $E_z$ must be spatially constant in a steady state, we have

$$\frac{\eta B_y}{l} \approx v_{in} B_y, \tag{3.2}$$

where the magnetic diffusivity $\eta \equiv \frac{c^2}{4\pi\sigma}$ for electrical conductivity $\sigma$ and we have approximated the current density by

$$J \sim \frac{cB_y}{4\pi l} .$$

Equations (3.1) and (3.2) together imply that

$$v_{in} = \left(\frac{v_A \eta}{L}\right)^{\frac{1}{2}} \equiv v_{PS} . \tag{3.3}$$

Defining the magnetic Reynolds number $R_m \equiv \frac{L v_A}{\eta}$, we have $v_{PS} = v_A R_m^{-\frac{1}{2}}$ and $l = L R_m^{-\frac{1}{2}} \equiv l_{PS}$.

In applying this reconnection model to colliding flux tubes, we identify $B_y$ with $B_\phi(R)$ and $\rho$ with its value near the edge of the tube. According to the results of Section 2, the dynamics of the collision between the flux tubes determines the relationship between the collision velocity, $R\frac{d\epsilon}{dt} \equiv \dot{\epsilon}$ and the length of the contact region, $L = (8\epsilon)^{\frac{1}{2}} R$. Therefore, it would appear that eqn. (3.3) can be satisfied only under special circumstances.

Suppose the incoming velocity $R\dot{\epsilon} < v_{PS}$. In that case, advection of the field by the incoming flow can be balanced by diffusion in a layer of width $l \approx l_{PS} \frac{v_{PS}}{R\dot{\epsilon}}$. On the other hand, if $R\dot{\epsilon} > v_{PS}$, the fieldlines are being brought in too rapidly to reconnect and magnetic flux piles up in the layer. Flux pileup has two effects: $v_A$ is increased, thereby increasing $v_{PS}$ (see eqn. (3.3)), and the incoming fluid is decelerated by the magnetic pressure gradient. The fluid velocity and $v_{PS}$ adjust until they balance, and thus we reach a state in which Parker-Sweet reconnection can take place. Note that flux pileup is not inconsistent with the assumption of incompressible flow, because the fluid can adjust its density by motion parallel to the common boundary without perturbing the magnetic field. We can expect this to occur as long as the tubes are gas pressure dominated or weakly twisted.

We quantify these ideas and derive the flux pileup behavior by using conservation of momentum to relate the magnetic field $B_\phi(R)$ at the outer edge of the deceleration layer to the magnetic field $B_{\phi l}$ at the inner edge of the deceleration region, which is also the edge of the reconnection region:

$$\rho(R\dot{\epsilon})^2 + \frac{B_\phi^2(R)}{8\pi} = \rho v_{in}^2 + \frac{B_{\phi l}^2}{8\pi}, \tag{3.4}$$

Dividing eqn (3.4) by $\rho$, substituting for $v_{in}$ using eqn. (3.3) with $v_A$ set equal to $v_{Al}$, the Alfven speed at the inner edge of the deceleration region, and introducing the Ohmic diffusion velocity

$$v_D \equiv \frac{\eta}{L} = \frac{v_A}{R_m}, \tag{3.5}$$

we derive a quadratic equation for $v_{Al}$ with solution

$$v_{Al} = (v_A^2 + 2(R\dot{\epsilon})^2 + v_D^2)^{\frac{1}{2}} - v_D. \tag{3.6}$$

If the collisions are slow and $R_m$ is large, the magnetic field is only slightly increased across the deceleration layer. To see this, recall that $R\dot{\epsilon} \leq \frac{1}{2}v_c$ and therefore is of order $M_{AC}v_A$. Using this result we see from eqn. (3.6) that

$$\frac{v_{Al}}{v_A} \leq (1 + \mathcal{O}(\mathrm{M}_{AC}^2) + \mathrm{R}_m^{-2})^{\frac{1}{2}} - \mathrm{R}_m^{-1}. \tag{3.7}$$

Equation (3.7) holds only if $\frac{B_{\phi l}}{B_\phi(R)} > 1$.

In summary, consider the history of a flux tube collision from first contact to maximum deformation. According to eqn. (2.9), the speed of each tube decreases monotonically from $\frac{1}{2}v_c$ to zero. Initially, $v_{PS}$ is large because the length of the common boundary, $L$, is small (see eqn. [3.3]), and fieldlines reconnect only through diffusion. But $v_{PS}$ drops as the contact surface lengthens, and so a time may come when $R\dot{\epsilon} > v_{PS}$. This marks the onset of the Parker-Sweet reconnection phase. This phase ends when $R\dot{\epsilon}$ drops below $v_{PS}$ once again, due to the deceleration of the flux tubes. At this point the system is not driven hard enough for Parker-Sweet reconnection to continue, and any residual reconnection occurs only through Ohmic diffusion.

If the collision is sufficiently weak, $R\dot{\epsilon}$ never exceeds $v_{PS}$, and the Parker-Sweet reconnection phase is never established. We can establish a criterion for the existence of a reconnection phase by substituting eqns. (2.1) and (2.9) into eqn. (3.3) to yield the condition that the tube velocity equals $v_{PS}$:

$$\epsilon^{\frac{1}{2}}(\epsilon_{max}^{\frac{5}{2}} - \epsilon^{\frac{5}{2}}) = \frac{\mu}{2^{\frac{3}{2}}\alpha R_{mg}}, \tag{3.8}$$

where $R_{mg} \equiv R_m(\frac{R}{L})$ is the global magnetic Reynolds number. There is a reconnection phase only when the equality (3.8) can be fulfilled, which is seen to be when

$$\epsilon_{max} > (\frac{6^{\frac{6}{5}}\mu}{2^{\frac{3}{2}}5\alpha R_{mg}})^{\frac{1}{3}} \equiv \epsilon_c. \tag{3.9}$$

Or, eliminating $\epsilon_{max}$ in favor of $M_{AC}$ using eqn.(2.10) and introducing a global Ohmic drift velocity

$$v_{Dg} \equiv \frac{v_A}{R_{mg}},$$

we find that Parker-Sweet reconnection occurs for

$$v_c > \frac{2^{\frac{7}{8}}3^{\frac{1}{2}}}{5^{\frac{5}{12}}}(\frac{\alpha}{\mu})^{\frac{1}{12}}v_A^{\frac{7}{12}}v_{Dg}^{\frac{5}{12}} \approx 1.65(\frac{\alpha}{\mu})^{\frac{1}{12}}v_A^{\frac{7}{12}}v_{Dg}^{\frac{5}{12}} \equiv v_{crit}. \tag{3.10}$$

Equations (3.10) shows that an approximate criterion for the existence of a Parker-Sweet reconnection phase is that the relative collision velocity exceeds the geometric mean of the global Ohmic diffusion speed and the azimuthal component of the Alfven speed. This result is almost independent of the unknown parameter $\alpha$ introduced in eqn. (2.8) and the parameter $\mu$ (introduced in eqn. (2.6)) which is determined by the various quantities characterizing the flux tubes and background medium. We expect $v_{Dg} \ll v_A$ in a highly conducting medium, so there is a wide range over which the collisions are rapid enough to force reconnection but slow enough to be described by the small $M_{AC}$ approximation.

### 3.2. The Amount of Reconnected Flux

We now calculate the amount of magnetic flux that is reconnected during the Parker-Sweet phase. Let the two positive, real roots of eqn. (3.8), which delimit the interval during which reconnection occurs, be $\epsilon_i$ and $\epsilon_f$, and let the corresponding times be $t_i$ and $t_f$. The reconnected flux per unit length in $z$ is

$$\Phi_R = \int_{t_i}^{t_f} dt B_{\phi l} v_{PS}, \tag{3.11}$$

where $B_{\phi l}$ and $v_{PS}$ are both functions of time and are determined by eqns. (3.3) and (3.4). Equation (3.7) shows that we can set $B_{\phi l} \approx B_\phi(R)$. Equation (3.11) can be rewritten as an integral over $\epsilon$ using eqns (2.1), (2.8), (2.9), and (3.3). We find

$$\frac{\Phi_R}{RB_\phi} = \frac{M_{AC}^{4/5}}{\alpha^{2/5}} y_0^{1/4} I(y_0), \tag{3.12}$$

where $y \equiv \frac{\epsilon}{\epsilon_{max}}$,

$$I(y_0) \equiv \int_{y_i}^{y_f} \frac{dy}{y^{\frac{1}{4}}(1-y^{\frac{5}{2}})^{\frac{1}{2}}}, \tag{3.13}$$

and the parameter $y_0$ is defined as

$$y_0 \equiv \frac{\mu^2 \alpha^{2/5}}{8R_{mg}^2 M_{AC}^{24/5}}$$

and $y_i$ and $y_f$ are the positive real roots of the equation

$$y^{\frac{1}{2}} - y^3 - y_0^{\frac{1}{2}} = 0, \tag{3.14}$$

which follows from dividing eqn. (3.8) by $\epsilon_{max}^3$. It follows from eqn. (3.9) that $y_0$ cannot exceed $y_{0,max} \equiv \frac{25}{6^{\frac{12}{5}}}$.

The integral $I(y_0)$ and the limits of integration $y_i$ and $y_f$ cannot be found analytically, but they can be found approximately in certain cases, and this is useful in developing a scaling for the reconnected flux. Near threshold, when $y_0$ is slightly less than $y_{0,max}$, we find

$$I(y_0) \approx \frac{4 \cdot 6^{\frac{9}{10}} \delta^{\frac{1}{2}}}{5^{\frac{3}{2}}}, \tag{3.15}$$

where $\delta \equiv y_{0,max} - y_0$. For tube velocities much greater than the threshold given by eqn. (3.10), $y_0$ approaches zero and $y_i \approx 0$, $y_f \approx 1$. In this case

$$I(y_0 = 0) \equiv I_{max} = \frac{2\pi^{\frac{1}{2}}\Gamma(\frac{3}{10})}{5\Gamma(\frac{4}{5})} \approx 1.822. \tag{3.16}$$

We have computed $\Phi_R$ numerically as well. Figure (3) is a plot of $y_0^{1/4} I(y_0)$ versus $y_0$. As eqn. (3.12) shows, this quantity is proportional to the reconnection rate. Note that there is a maximum value of $\Phi_R$: when $M_{AC}$ is near its threshold value for reconnection $\Phi_R$ increases with increasing Mach number, primarily because the interval during which reconnection can take place is increasing. As $M_{AC}$ increases further, reconnection can take place at almost all times in the collision, but the collision itself takes less time (the collision time scales as $M_{AC}^{-\frac{1}{5}}$), so $\Phi_R$ begins to decrease again.

## 4. THE CRITICAL VELOCITY FOR MERGING

We follow B84 in estimating the critical velocity for merging of the tubes: if the magnetic energy per unit length of the reconnected flux exceeds the kinetic energy per unit length of the colliding tubes, the tubes coalesce.

The magnetic energy $W_M$ can be written in terms of an effective length $\Delta R_{eff}$ defined by

$$\Delta R_{eff} \equiv \frac{\Phi_R}{B_\phi(R)}.$$

It is important for the validity of our theory that $\frac{\Delta R_{eff}}{R\epsilon_{max}}$ be small, because we have solved for the structure of the reconnection layer only in an approximate way. It can be shown that this ratio is just $I(y_0) y_0^{\frac{1}{4}}$, which is always less than 1 (see Figure 3). That is, only a small fraction of the fieldlines in the deformation layer are reconnected.

The magnetic energy of the reconnected flux is

$$W_M = \frac{B_\phi^2}{8\pi} 2\pi R \Delta R_{eff} = \frac{RB_\phi(R)\Phi_R}{4}. \tag{4.1}$$

The kinetic energy $K$ per unit length of each tube is

$$K = \frac{\mathcal{M}_{eff} v_c^2}{8} = \frac{M_{AC}^2 R^2 B_\phi^2(R)}{8}. \tag{4.2}$$

Using eqns (3.12), (4.1), and (4.2) we have

$$\frac{W_M}{K} = \frac{2^{\frac{7}{4}}}{\mu^{\frac{1}{2}}\alpha^{\frac{1}{2}}} R_{mg}^{\frac{1}{2}} y_0^{\frac{1}{2}} I(y_0), \qquad (4.3)$$

with the criterion for coalescence $\frac{W_M}{K} > 1$.

The function $y_0^{1/2} I(y_0)$ vanishes at both ends of its range $(0, y_{0,max})$, and attains a maximum value of about 0.35 at $y_0 \approx 0.13$. As $R_{mg}$ is expected to be large, $W_M/K$ certainly exceeds unity at its maximum, so coalescence occurs over an interval in the collision parameter $y_0$ with endpoints given approximately by the roots of the equation

$$y_0^{1/2} I(y_0) = \left(\frac{\alpha\mu}{2^{7/2} R_{mg}}\right)^{1/2}. \qquad (4.4)$$

We solve eqn. (4.4) by approximating $I(y_0)$ near the ends of its range using eqns (3.15) and (3.16). Near $y_0 = 0$, $I(y_0) \approx I_{max}$, and we find that coalescence occurs for

$$y_0 > \frac{\alpha\mu}{I_{max}^2 2^{7/2} R_{mg}}, \qquad (4.5)$$

or

$$M_{AC} < \left(\frac{2^{1/2} I_{max}^2 \mu}{\alpha^{3/5} R_{mg}}\right)^{5/24}. \qquad (4.6)$$

The smallness of the lower limit of $y_0$ justifies the approximation $I \approx I_{max}$. It can be shown froms eqns. (3.10) and (3.15) that the larger root of eqn. (4.4) is nearly $y_{0,max}$ itself, corresponding to the minimum velocity for reconnection, $v_{crit}$, given in eqn. (3.10). Putting these results together and using eqn. (2.10) we see that coalescence is expected for collision Mach numbers in the range $M_{AC,min}$ to $M_{AC,max}$, where

$$M_{AC,min} = \frac{3^{1/2}\alpha^{1/12}\mu^{5/12}}{5^{5/12} 2^{1/8} R_{mg}^{5/12}}, \qquad (4.7)$$

and

$$M_{AC,max} = \left(\frac{2^{1/2} I_{max}^2 \mu}{\alpha^{3/5} R_{mg}}\right)^{5/24} = 1.69 M_{AC,min}\left(\frac{R_{mg}}{\alpha\mu}\right)^{5/24}. \qquad (4.8)$$

We will see in the next section that the range of Alfven Mach numbers for which coalescence occurs is between one and two orders of magnitude.

# 5. DISCUSSION

In this section we critically review the main assumptions underlying our calculation. We then evaluate some of the formulae for parameters appropriate to the solar convection zone and discuss flux tube merging in that environment. Finally, we summarize the main results of the paper.

## 5.1. Validity of Analysis

We assumed in this paper that magnetic fluxtubes are thin and straight and have twisted field lines. The first two properties appear to be true of the tubes directly observed in the solar photosphere. Severely kinked or bent tubes are subject to strong magnetic tension restoring forces, so the assumption of straightness is probably a reasonably good one. There is no direct observational evidence that the field lines twist. However, magnetic fields which are concentrated by three dimensional flows possessing vorticity, as is the case in stellar convection zones, might well retain twist in addition to an axial component. Tubes which originate in the overshoot region at the base of the convection zone may also possess twist (Cattaneo, Chiueh, & Hughes 1990). The assumption of twist is central to the collision model.

Following B84 , we considered direct collisions between antiparallel tubes. This is a maximally symmetric situation, and, as such, a singular one. Although extending our results to more general flux tube orientations is beyond the scope of this paper, the reconnection process might not be too different if the tubes are not perfectly aligned, in which case some combination of the axial and azimuthal components would reconnect.

The collision model assumes incompressible flows and slow collisions. If the relative velocities are comparable to the convective velocities deep in a stellar interior, and if the tubes are gas pressure dominated, the collisions are in fact likely to be slow. If the tubes are well coupled by friction to the surrounding gas, then the typical relative velocities of colliding tubes will in fact be much less than the mean convective eddy velocity. On the other hand, rapid relative motion is possible even with frictional coupling if the buoyancy and/or tension forces are strong enough (e.g. Moreno-Insertis 1986). Our conclusion that the amount of flux reconnected is a decreasing function of tube speed may change if we relax the assumptions that the collisions are slow, that the tubes are not greatly deformed, and that there are no important dissipation mechanisms other than resistivity. For instance, if a collision is strong enough to drive shocks in the tubes, there would be substantial viscous dissipation and the tubes might stay together long enough for extensive reconnection to occur. Likewise, tubes could be disrupted by violent, oblique collisions. However, for slow collisions, our flux pileup model should be generalizable to kinematic models other than B84 's.

Finally, our description of the reconnection region is approximate, and based on the Parker-Sweet theory. We ignored additional field annihilation resulting from Ohmic diffusion. However, the contribution of the latter is expected to be very small.

## 5.2. Application to the Solar Convection Zone

In this section we apply our theory to the solar convection zone. Except where noted otherwise, our numerical values are derived from the convection zone model of Spruit

(1977).

The global magnetic Reynolds number $R_{mg}$ scales as $\frac{RB_\phi}{\eta\rho^{\frac{1}{2}}}$. The denominator of this quantity is roughly constant throughout the convection zone and approximately equal to $10^4$. Measuring $R$ in units of $10^2 km$ and $B_\phi$ in units of $10^2 G$ we find

$$R_{mg} \approx 10^5 R_2 B_2. \tag{5.1}$$

Note that the axial fieldstrength in the tubes is likely to be higher than $B_\phi$; therefore $R_{mg}$ computed using the entire magnetic fieldstrength is significantly larger.

We next compute the lower and upper limits $v_{cl}$ and $v_{cu}$ to $v_c$ for which our model predicts merging. The lower limit is given by eqn. (3.10) and using eqn. (5.1) it is given by

$$v_{cl} \approx .34 (\frac{\alpha B_2^7}{\mu R_2^5})^{\frac{1}{12}} (\rho)^{-\frac{1}{2}} cm s^{-1}. \tag{5.2}$$

The upper limit $v_{cu}$ follows from eqns. (2.6) and (4.4), and is given by

$$v_{cu} \approx \frac{6.4}{\rho^{\frac{1}{2}}} (\frac{B_2^{19}}{\alpha^3 \mu^{12} R_2^5})^{\frac{1}{24}}. \tag{5.3}$$

The ratio of these two speeds is given by

$$\frac{v_{cu}}{v_{cl}} \approx 18.8 (\frac{B_2 R_2}{\mu^2 \alpha})^{\frac{5}{24}}. \tag{5.4}$$

We see that the range of speeds for which our theory predicts coalescence is rather small; generally between one and two orders of magnitude.

We now estimate $v_{cl}$ and $v_{cu}$ for parameters appropriate to the solar convection zone, and compare them with the characteristic fluid velocities $v_{conv}$ expected from mixing length theory. A range of velocities on different size scales is undoubtedly present in the convection zone, and as stated in the preceding section, the degree to which the tubes are coupled to the fluid motion is quite uncertain. Therefore, the mixing length velocity is only a rough benchmark.

Halfway to the bottom of the convection zone the density is $\rho \approx 6.4 \times 10^{-2} gm\ cm^{-3}$. This leads to values of $v_{cl}$ and $v_{cu}$ of 1.4 and 26 $cm/s$, respectively, with all parameters in eqns. (5.2) and (5.3) having their fiducial values. On the other hand, $v_{conv} \approx 9.0 \times 10^3 cm/s$. The magnetic fieldstrength $B_{eq}$ at which the kinetic and magnetic energies are equal at this depth is $\approx 8000G$; if we assume arbitrarily that $B_\phi \approx 0.1 B_{eq}$, $v_{cl}$ is boosted to 4.8 $cm/s$ and $v_{cu}$ to 137 $cm/s$. We interpret the lower limit as essentially zero, while the upper limit is still a small fraction, no more than 1.5% in this example, of the convective velocity.

At the base of the convection zone $\rho \approx 0.23 gm\ cm^{-3}$, so $v_{cl}$ and $v_{cu}$ are reduced to .67 and 12 $cm/s$, respectively. At this depth, $v_{conv}$ is only 86 $cm/s$. Increasing the magnetic fieldstrength to $0.1 B_{eq}$ makes little difference in this case, as $B_{eq}$ is only $1460G$; $v_{cu}$ would then be 16 $cm/s$. Small as this value is it is about 18% of the convective velocity, so a fairly large fraction of flux tube collisions could result in coalescence by the processes included in our theory.

Near the top of the convection zone, at a depth of $2 \times 10^9 cm$, $\rho \approx 3.47 \times 10^{-4} gm$ $cm^{-3}$. Then $v_{cl}$ and $v_{cu}$ are about 6 and 110 $cm/s$; again small compared with the estimated $v_{conv}$ of $2.1 \times 10^4 cm/s$. At this depth $B_{eq} \approx 4400 G$, so assuming $B_2 \sim 4.4$ increases $v_{cl}$ to 14 $cm\ s^{-1}$ and $v_{cu}$ to 360 $cm/s$, less than 1% of $v_{conv}$.

In summary, if the relative velocities between flux tubes are on the order of the convective velocity as obtained from mixing length theory, only a small fraction of collisions are slow enough to allow coalescence by reconnection. This fraction appears to increase monotonically with depth in the convection zone.

Notice that the implied magnetic fluxes in our fiducial tubes are rather small; if the axial component of $B$ is about a kilogauss, the flux is approximately that contained in a small photospheric flux tube. If the radii of flux tubes in the convection zone exceed 100 km, eqns. (5.2) and (5.3) show that $v_{cl}$ and $v_{cu}$ are both reduced, although their dependences on $R$ are rather weak. Stated in another way, if the tubes carry more axial flux (at constant axial fieldstrength) then $v_{cl}$ and $v_{cu}$ are both reduced, although their ratio is increased by $R_{mg}^{5/24}$.

### 5.3 Summary

In this paper we have considered magnetic reconnection during slow collisions between twisted magnetic flux tubes. Our underlying assumption is that the tubes behave as described by Parker (1982), Moreno-Insertis (1986), and others: they are relatively long-lived, coherent structures and their motion is determined by frictional coupling to the background fluid, magnetic tension, and buoyancy. The frictional coupling may be strong but our model requires that it not be perfect, otherwise the dynamics of tube collisions would be determined entirely by the ambient flow and not at all by internal forces resulting from deformation of the tubes. Thus, for example, two tubes might intersect at an oblique angle and be dragged across each other: if magnetic tension is unable to snap them back and straighten them out, some forced reconnection is probably inevitable. This may occur in some situations. However, the analysis in this paper is restricted to flux tubes which are kept fairly straight (because of the flow, because of magnetic tension forces, or both) and which move semi-independently of the surrounding fluid. The tubes may be brought together because they are advected with the flow, through self-generated hydrodynamical forces (Parker 1979b), or some combination of them. Once the tubes are very close, the background flow must be essentially constant throughout the interaction region, so the dynamics of the collision should be governed entirely by internal forces. Although the overpressure associated with the deformation of colliding tubes tends to separate them, a sufficient amount of reconnected azimuthal magnetic flux will form a collar that keeps the tubes together. If the collision is too weak, there is so little deformation of the tubes that very little reconnection occurs, and therefore the tubes do not coalesce. If the collision is too strong it is so rapid that reconnection has very little time to occur, and again there is no coalescence. This is our basic picture, which we flesh out in the following paragraphs.

We showed that Parker-Sweet reconnection of the magnetic field can occur at the interaction surface provided that the collision velocity exceeds a threshold which is ap-proximately the geometric mean of the global Ohmic diffusion velocity in the tube and the Alfven crossing time of the tube, (measured with respect to the azimuthal field component only). The reconnection occurs through a very small amount of flux pileup at the edge of the interaction region, and only a small fraction of the flux in the interaction region is reconnected. There is a velocity which maximizes the amount of reconnected flux; it is set by balancing the driving, which increases with collision velocity, against the decrease in collision duration with increasing velocity, which reduces the time during which reconnection can occur. If the collision velocity is less than the threshold, the fields are not driven hard enough to reconnect, but they will diffuse. Our results on reconnection are relatively insensitive to the values of the two parameters $\alpha$ and $\mu$ (introduced in eqns. (2.6) and (2.8) respectively) which characterize the collision but are not well determined by the flux tubes model. We also argued in Section 5.1 that the essential features of this picture are preserved even if the tubes are not perfectly aligned. If the collisional velocity approaches the Alfven speed, our theory breaks down in many places, starting with the dynamics of the collision.

We then discussed tube coalescence, arguing that merging occurs if the energy in the reconnected field exceeds the relative energy of the collision. This led to the result that merging occurs only for collision velocities within the range given by eqns. (5.2) and (5.3). We compared this velocity range with solar convection velocities estimated from mixing length theory, and found that they are always substantially smaller- although the gap narrows with depth in the convection zone. On this basis we argue that collisions between flux tubes do not generally lead to coalescence, although they might disrupt the flux tube edges. Again, our results are quite insensitive to $\alpha$ and $\mu$. The collision problem at high Mach number introduces effects we have not included, such as severe deformation and shock waves, and is probably amenable only to numerical analysis.

**Acknowledgements:** We wish to thank Tom Bogdan and Brad Hindman for useful discussions. JER's work was supported in part by the Princeton University Department of Astrophysical Sciences. EGZ's work was supported by NSF grant ATM–901275 and the NASA Space Physics Theory Program.

## APPENDIX

Our efforts to understand the restoring force in a collision of two parallel twisted magnetic flux tubes suggested that we investigate an analogous collision of two elastic solid cylinders. The usual heuristic comparison of Alfven waves traveling along a magnetic field line and mechanical waves on a string also leads to the comparison of MHD with the theory of elasticity.

We assume an infinitely conducting magnetized fluid and consider the reaction of the system to a small perturbation away from an equilibrium state. The standard theory of elastic solids is likewise based on linearized equations describing the response of a system to a small perturbation from equilibrium. The equilibrium state is a minimum of the free energy, so the first nonvanishing term in the Taylor expansion of the free energy is generically of quadratic order. This leads to Hooke's law for an elastic solid and the usual linearized MHD equations for a magnetized fluid. We expect to need the equations of anisotropic elasticity, since the magnetic field is inherently anisotropic. Moreover, magnetic fields possess directionality, a feature that is absent even in anisotropic elastic solids. This difference is expected to limit the ultimate utility of the analogy.

The most general form of Hooke's law, applicable for anisotropic elastic solids (e.g., crystals), is

$$\rho \ddot{\xi}_i = \lambda_{iklm} \partial_k \partial_l \xi_m \tag{A1}$$

(Cf. Landau and Lifschitz 1970.) Here $\vec{\xi}$ is the displacement of a volume element from its equilibrium position, which is assumed to be small, $\lambda$ is the elastic modulus tensor, $\rho$ is the density, $\partial_j$ denotes partial differentiation with respect to spatial coordinate $j$, $\dot{}$ denotes differentiation with respect to time, and summation over repeated indices is implied. The elastic modulus tensor must satisfy the symmetries

$$\lambda_{iklm} = \lambda_{ikml} = \lambda_{kilm} = \lambda_{lmik} \tag{A2}$$

(Landau and Lifschitz 1970).

We make a local approximation and assume a a uniform fluid at rest and a constant background magnetic field. The linearized magnetic induction equation can be integrated once in time to give the perturbed magnetic field $\delta \vec{B}$ in terms of the fluid displacement $\vec{\xi}$ in coordinate notation as

$$\delta B_j = B_k \partial_k \xi_j - B_j \partial_k \xi_k \ . \tag{A3}$$

The linearized equation of motion for $\vec{\xi}$ is

$$4\pi \rho \ddot{\xi}_i = -\partial_i (B_k \delta B_k) + B_k \partial_k \delta B_i = B_k (\partial_k \delta B_i - \partial_i \delta B_k) \ . \tag{A4}$$

We substitute $\delta \vec{B}$ as given by eqn. (A3) into eqn. (A4) to obtain

$$4\pi \rho \ddot{\xi}_i = -B_j B_k \partial_i \partial_k \xi_j + B^2 \partial_i \partial_l \xi_l + B_j B_k \partial_j \partial_k \xi_i - B_i B_j \partial_j \partial_l \xi_l \ . \tag{A5}$$

We can rewrite this in a form resembling Hooke's law,

$$4\pi \rho \ddot{\xi}_i = +B^2 \delta_{ik} \delta_{lm} \partial_k \partial_l \xi_m - \delta_{ik} B_l B_m \partial_k \partial_l \xi_m - B_i B_k \delta_{lm} \partial_k \partial_l \xi_m + \delta_{im} B_k B_l \partial_k \partial_l \xi_m \ . \tag{A6}$$

We then define

$$4\pi \lambda^{(1)}_{iklm} := B^2 \delta_{ik} \delta_{lm}$$
$$4\pi \lambda^{(2)}_{iklm} := -\delta_{ik} B_l B_m - B_i B_k \delta_{lm}$$
$$4\pi \hat{\lambda}^{(3)}_{iklm} := \delta_{im} B_k B_l$$

so that $\lambda^{(1)}$ and $\lambda^{(2)}$ are tensors with the symmetry properties of an elastic modulus tensor but the tensor $\hat{\lambda}^{(3)}$ lacks these properties. We note in passing that by orienting a coordinate axis along the uniform background magnetic field we reduce the number of nonvanishing components of the $\lambda^{(j)}$ well below the maximum of 21 (cf. Landau and Lifschitz) allowed by the symmetry properties of the elastic modulus tensor.

We now examine $\hat{\lambda}^{(3)}$ more closely. We can try to symmetrize it, defining

$$\tilde{\lambda}^{(3)}_{iklm} := (\hat{\lambda}^{(3)}_{iklm} + \hat{\lambda}^{(3)}_{kilm} + \hat{\lambda}^{(3)}_{ikml} + \hat{\lambda}^{(3)}_{lmik})/4$$
$$= \pi(\delta_{im} B_l B_k + \delta_{km} B_l B_i + \delta_{il} B_m B_k + \delta_{kl} B_i B_m)$$

We can now isolate the non-elastic part of the response, writing

$$\rho \ddot{\xi}_i = (\lambda^{(1)} + 2\lambda^{(2)} + 4\tilde{\lambda}^{(3)})_{iklm} \partial_k \partial_l \xi_m - B_i B_m \partial_k \partial_k \xi_m \quad . \tag{A9}$$

If the last term on the right hand side is much smaller than the left hand side (i.e., small to the same order as other terms omitted in the derivation), then we have an approximate elastic modulus tensor describing the response of an MHD system to perturbations. In fact, we recognize that this term contains the scalar product of $\vec{\xi}$ with $\vec{B}$. One can show from the original equation of motion (A4) that this vanishes. Therefore, small amplitude motion is confined to planes perpendicular to $\vec{B}$ and is described by an elasticity tensor.